\documentclass[10pt,conference]{IEEEtran}
\usepackage{epsfig,setspace,amsmath,epsf,amssymb,bm,theorem,cite,graphicx, epstopdf,algorithm,algpseudocode,float,color,mathtools,authblk}
\usepackage[table,xcdraw]{xcolor}
\usepackage{booktabs}
\usepackage{subfigure}
\usepackage{bbm}

\newtheorem{lemma}{Lemma}

\newenvironment{Proof}[1]{\medskip\par\noindent{\bf Proof:\,}\,#1}{{\mbox{\,$\blacksquare$}\par}}

\IEEEoverridecommandlockouts
\allowdisplaybreaks
\begin{document}

\title{Timely Tracking of a Wiener Process With \\ Single Bit Quantization}

\author{Ismail Cosandal \qquad Sahan Liyanaarachchi \qquad Sennur Ulukus\\
	\normalsize Department of Electrical and Computer Engineering\\
	\normalsize University of Maryland, College Park, MD 20742 \\
	\normalsize \emph{ismailc@umd.edu} \qquad \emph{sahanl@umd.edu} \qquad \emph{ulukus@umd.edu}}
\maketitle

\begin{abstract}
We consider the problem of timely tracking of a Wiener process via an energy-conserving sensor by utilizing a single bit quantization strategy under periodic sampling. Contrary to conventional single bit quantizers which only utilize the transmitted bit to convey information, in our codebook, we use an additional `$\emptyset$' symbol to encode the event of \emph{not transmitting}. Thus, our quantization functions are composed of three decision regions as opposed to the conventional two regions. First, we propose an optimum quantization method in which the optimum quantization functions are obtained by tracking the distributions of the quantization error. However, this method requires a high computational cost and might not be applicable for energy-conserving sensors. Thus, we propose two additional low complexity methods. In the \emph{last-bit aware} method, three predefined quantization functions are available to both devices, and they switch the quantization function based on the last transmitted bit. With the \emph{Gaussian approximation} method, we calculate a single quantization function by assuming that the quantization error can be approximated as Gaussian. While previous methods require a constant delay assumption, this method also works for random delay. We observe that all three methods perform similarly in terms of mean-squared error and transmission cost.
\end{abstract}

\section{Introduction}
\begin{figure}
    \centering
    \includegraphics[width=0.95\linewidth]{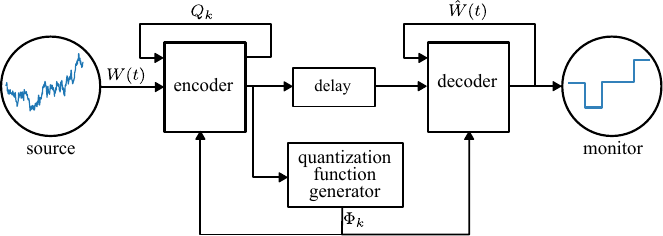}
    \caption{A general system model for timely tracking of a Wiener process. The quantization function generator updates the quantization functions based on the transmitted bit, and the new quantization function is available for both the encoder and the decoder.}
    \label{fig:sys}
\end{figure}

We consider the problem of timely tracking a Wiener process (a.k.a. zero-drift Brownian motion) across a delay channel by enforcing a periodic sampling policy along with a single bit quantization scheme. In each period, the source encodes the difference between the source process and the estimation on the monitor, which is equivalent to the sum of the previous quantization error and the increment of the process from the previous sampling time. In the most general case, the quantization function is subject to change based on the previously transmitted bit, and that quantization function is available to both the source and the monitor. To minimize the mean squared error (MSE), the monitor applies optimum decoding after the bit is transmitted; see Fig.~\ref{fig:sys}.

Recently, the timely tracking problem of a Wiener process has been investigated in the literature with different settings \cite{sun2017remote, Wei_chen, tsi_wei_chen, liyanaarachchi2025source}. A pioneering work \cite{sun2017remote} considers the lossless transmission of the Wiener process, where they propose \emph{signal-dependent} and \emph{signal-independent} sampling methods. One observation of this paper is that the MSE of signal-independent sampling is equivalent to the age of information (AoI) for the same sampling method. They additionally propose a signal-dependent sampling policy, in which the process is sampled after a threshold is exceeded, and it is shown that this method outperforms the signal-independent one.

In \cite{Wei_chen}, a similar framework has been investigated for lossy sampling with high-rate quantization. In \cite{Wei_chen}, the aforementioned analogy between MSE and AoI has been extended for the quantization error, and it has been shown that, in lossy sampling, MSE is equivalent to AoI with an additional term for quantization error. Additionally, only the increment of the process between samplings is quantized, thus, the quantization error accumulates over time. To deal with this, they propose a multi-level error correction scheme that periodically quantizes the accumulated error from the previous level alongside the increment of the process. The transmission time of each transmitted bit is proportional to the number of bits used for quantization. This framework is later extended by considering an additional channel delay in \cite{tsi_wei_chen}.

A recent study \cite{liyanaarachchi2025source} considers a lossless transmission scheme with dynamic thresholds and the generate-at-will approach. The increment of the process is sampled whenever it reaches one of the thresholds, and the monitor receives the sampled process after a deterministic delay based on how many bits are used for the corresponding thresholds. The paper investigates a source coding design that minimizes MSE under the assumption that the transmission of each bit lasts for a known constant delay.

Mathematical limitations and properties of the optimum quantization are mostly available for high rates and general distributions \cite{cover1999elements, gersho2012vector}, because of the intractable nature of the optimum quantization scheme. However, transmitting multiple bits for each symbol causes longer transmission times, hence, larger MSE \cite{liyanaarachchi2025source, Wei_chen} for timely tracking problems. In addition, they require more complex encoding and decoding procedures, which may not be suitable for simple energy-conserving sensors. On the other hand, some studies investigate specific cases and distributions. For instance, the asymptotic behavior as the rate goes to zero for a Gaussian source is studied for different distortion metrics in \cite{marco2006low, marco2007low}. In our work, we also investigate a special case of the quantization problem with a single bit. From the nature of the Wiener process, we quantize a Gaussian random variable with an additional term that comes from the previous quantization error. 

We consider a codebook $\mathcal{C}=\{0,1,\emptyset\}$ where $\emptyset$ corresponds to not transmitting anything, thus, the quantization function divides the input distribution into three regions. We show that the optimum quantization functions can be obtained by Lloyd-Max algorithm \cite{gersho2012vector}. With these quantization functions, we reach an expected distortion that is very close to the expected distortion under the assumption that the quantization error is Gaussian. However, the computational complexity of the optimum quantizer is impractical, especially for simple sensors. Thus, we propose two additional quantization methods whose expected distortions are also close to the optimum one. First, we propose a \emph{last-bit aware} quantization method, where there are three predefined quantization functions and the encoder and decoder switch the quantization function they apply based only on the last transmitted bit. The second quantization method we consider is a static quantization function that is obtained by Gaussian approximation to the quantization error, which shows robust performance under random delay, thanks to its unbiased handling of the $\emptyset$ symbol.

\section{Single bit Periodic Sampling}
A Wiener process $W(t)$ is a continuous-time stochastic process defined to start from $W(0)=0$, at each infinitesimal time it evolves with an independent Gaussian distribution as
\begin{align}
    W(t+\xi)-W(t)\sim \mathcal{N}(0,\sigma^2\xi),
\end{align}
where $\mathcal{N}(0,\sigma^2)$ is the zero-mean Gaussian distribution with a variance of $\sigma^2$. The sensor samples the process $W(t)$ with a period of $T$, and quantizes it with a single bit quantization function $\Phi_k$ at time $kT$. The monitor obtains the quantized sample after a delay $d$ and it updates its estimation process $\hat{W}(t)$ after decoding it with the same quantization function.

For a sample $k$ taken at time $kT$, we consider a quantization function $\Phi_k$ with the codebook $\mathcal{C}=\{-,\emptyset,+\}$, where $\Phi_k: \mathbb{R} \to \{c_{k-},c_{k\emptyset},c_{k+}\}$ has three regions $\mathcal{R}_{k-}=(-\infty,\tau_{k-}]$, $\mathcal{R}_{k\emptyset}=(\tau_{k-},\tau_{k+})$, $\mathcal{R}_{k+}=[\tau_{k+},\infty)$ which map the input to three center points $c_{k-}$, $c_{k\emptyset}$, $c_{k+}$, respectively. If the input is in region $\mathcal{R}_{k-}$ or $\mathcal{R}_{k+}$ bits $0$ or $1$ is transmitted correspondingly for the symbols $-$ and $+$. Otherwise, if the input is in region $\mathcal{R}_{k\emptyset}$, nothing is transmitted, i.e., what we call the \emph{empty bit} is transmitted for the symbol $\emptyset$ without any transmission cost. For the distortion metric, we use MSE, thus the expected distortion equals the second moment of the quantization error $Q_k=W(t)-\Phi_k(W(t))$.

\begin{figure}
    \begin{center}
    \subfigure[]{\includegraphics[width=0.85\columnwidth]{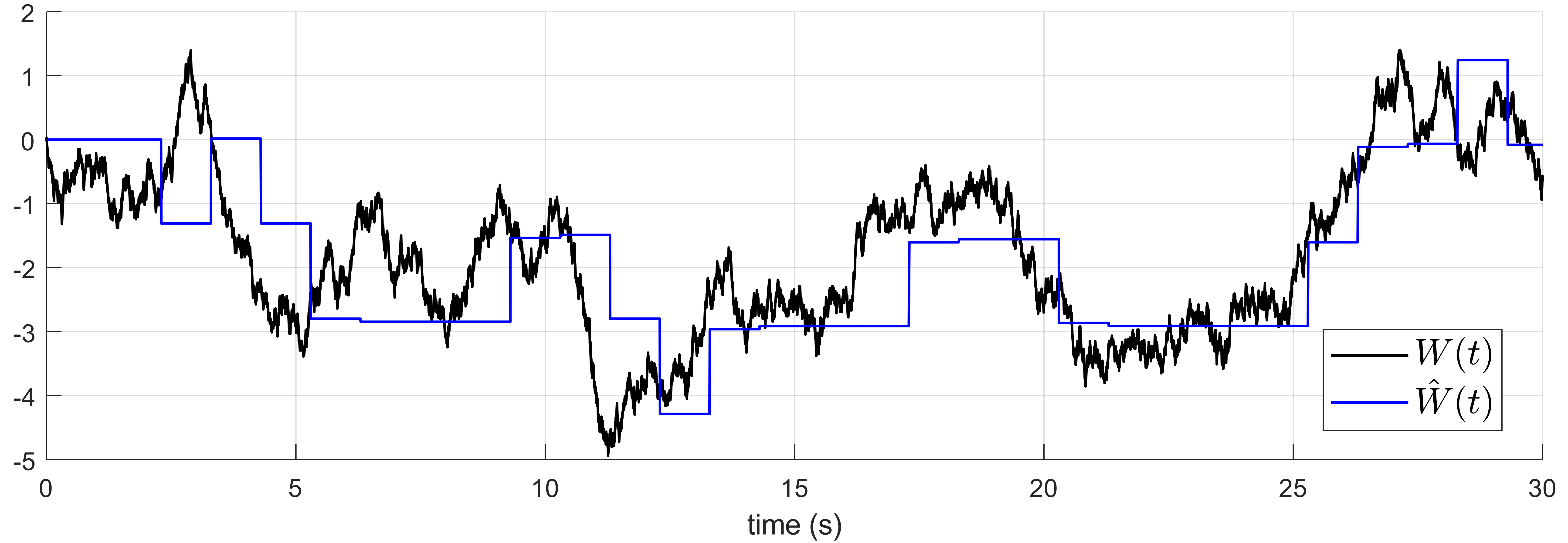}}
    \subfigure[]{\includegraphics[scale=0.55]{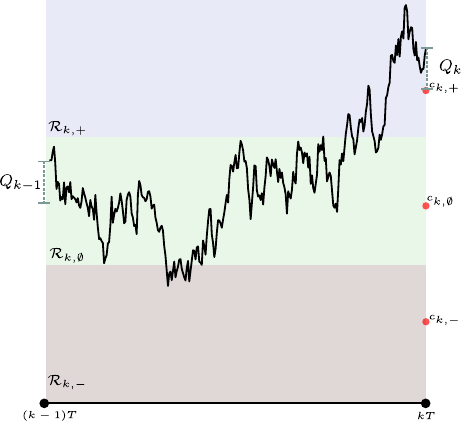}} ~ 
    \subfigure[]{\includegraphics[scale=0.55]{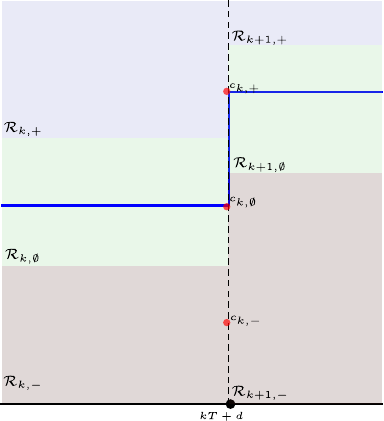}}  
    \end{center}  
    \caption{a) A sample path for $W(t)$ and $\hat{W}(t)$ for $\sigma^2=1$ and $d=0.3$. Quantization parameters for b) source and c) monitor within a single period.}
    \label{fig:example}
\end{figure}

The estimation processes is initiated as $\hat{W}(t)=0$. In the first transmission, the variable $Y_1=W(T)$ is quantized, and the monitor updates its estimate upon receiving and decoding the received bit as $\hat{W}(T+d)=\Phi_1\left(W(t)\right)$. From the Markov property of the Wiener process, this is the optimum estimation \cite{Wei_chen}, and the monitor keeps this estimation until a new update arrives. For the first period, the quantization error is $Q_1=W(T)-\Phi_1(W(T))$, and it is only available to the source.

For the remaining periods, the source quantizes the difference between the process and the current estimation. For the $k$th sample, the quantized variable can be expressed as
\begin{align}
    Y_k&=W({kT})-\hat{W}({kT})\\
    &=W({kT})-\left({W}\left({(k-1)T}\right)-Q_{k-1}\right)\\
    &=X_k+Q_{k-1},
\end{align}
where $X_k$ is the increment of the Wiener process between $(k-1)T$ and $kT$, which is independent and identically distributed with $\mathcal{N}(0,\sigma^2T)$. The transmitted bit is received by the monitor after a delay $d$, and the monitor updates its estimation accordingly as
\begin{align}
    \hat{W}({kT+d})=\hat{W}({kT+d'})+\Phi_k(Y_k), \label{eq:upd}
\end{align}
where $d'$ is the infinitesimal time before $d$, and the monitor keeps its estimation until receiving another bit. An example of how processes $W(t)$, $\hat{W}(t)$ and $\Phi_k$ evolve with time is illustrated in Fig.~\ref{fig:example}.

For now, we assume that the delay $d<T$ is constant, thus the monitor knows exactly when it is supposed to receive an update. Therefore, the monitor is able to update its estimation at $kT+d$ with $c_{\emptyset}$ even though nothing is transmitted, in other words, when an empty bit is transmitted at $kT$. The extension to random delays will be discussed in Section~\ref{sec:results}.

\begin{figure}[t]
    \centering
    \includegraphics[width=0.95\linewidth]{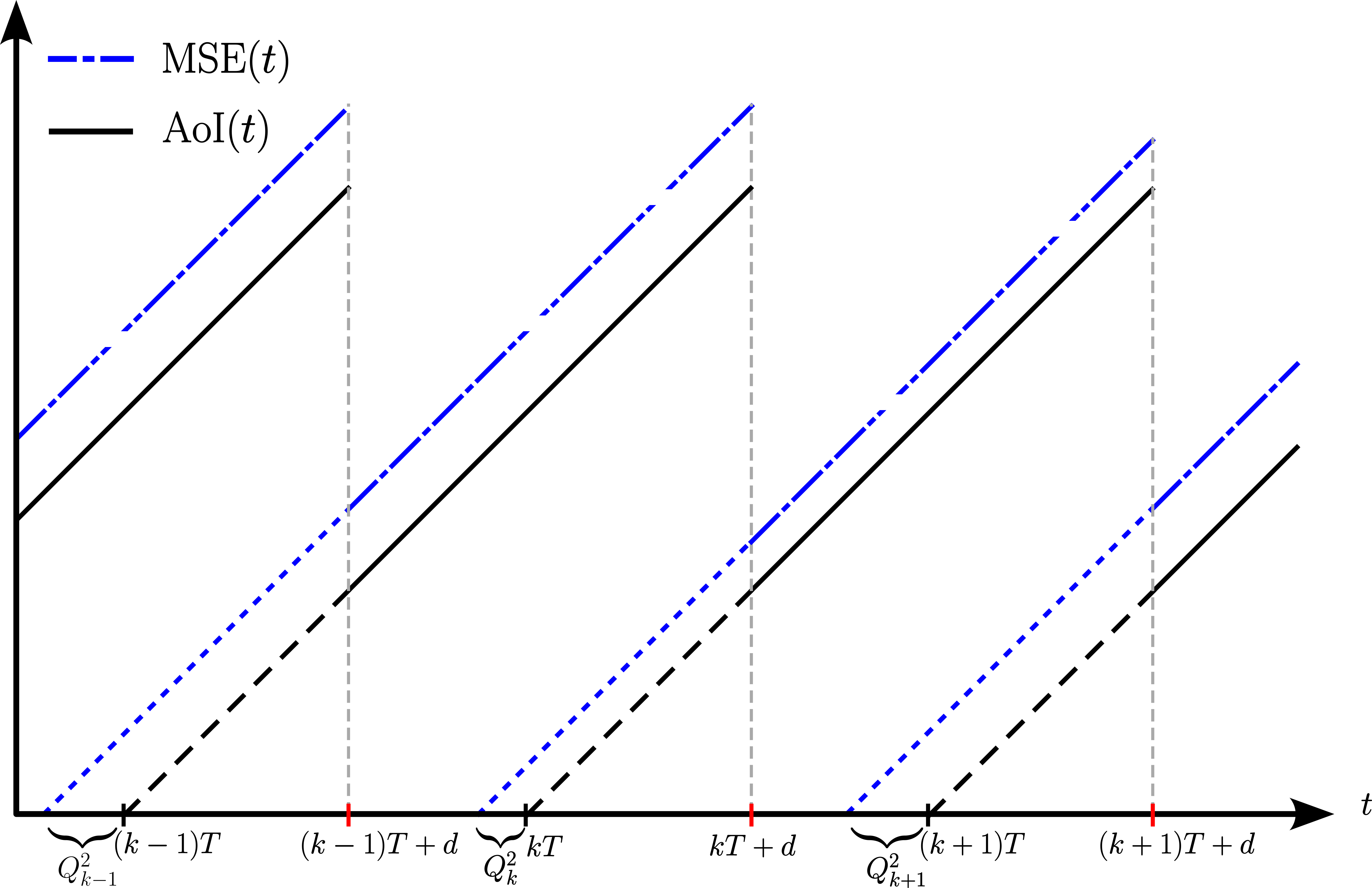}
    \caption{The analogy between MSE$(t)$ and AoI$(t)$.}
    \label{fig:aoi_mse}
\end{figure}

We aim to minimize the MSE between the processes $W(t)$ and $\hat{W}(t)$. The analogy between MSE and AoI has been highlighted in \cite{sun2017remote, Wei_chen}, and it can be adapted for our problem setting as follows.

\begin{lemma} \label{lem:2}
If the $k$th sample taken at time $kT$ has a quantization error $Q_k$ for a normalized Wiener process $(\sigma^2=1)$, the MSE$(t)$ at time $t$ is equivalent to the AoI as if the sample was taken at $kT-Q_k^2$ instead of $kT$. In essence, this is as if the quantization error induces an additional delay to the AoI process, as illustrated in Fig.~\ref{fig:aoi_mse}.
\end{lemma}

\begin{Proof}
The AoI is a measure of how long ago the latest received sample is taken \cite{Yates__HowOftenShouldone, yates2020age}, 
\begin{align}
    \text{AoI}(t)=t-k_tT,
\end{align}
where $k_t=\underset{k}{\arg\max}(kT+d<t)$.
From the update rule of the estimation in \eqref{eq:upd}, $\hat{W}(t)$ can be expressed as
\begin{align}
    \hat{W}(t)=W(k_tT)+Q_{k_t}.
\end{align}
Then, MSE$(t)$ can be calculated as
\begin{align}
    \text{MSE}(t)&= \mathbb{E} \left[| W(t)-\hat{W}(t) |^2\right] & \\
      &= \mathbb{E} \left[| W(t)-(W(k_tT)-Q_{k_t}) |^2\right] \\
    &= \mathbb{E} \left[| W(t-k_tT) |^2\Big|W(0)=Q_{k_t}\right]\\
    &=t-(k_tT-Q_{k_t}^2). \label{eq:mse}
\end{align}
\end{Proof}
    
This result can be generalized for any $\sigma^2$ as
\begin{align}
    \text{MSE}(t)=\sigma^2(t-k_tT)+Q_{k_t}^2.\label{eq:mse_gen}
\end{align}

From this result, we can define the expected value of the average MSE in the time interval $\begin{bmatrix}
    (k-1)T+d & kT+d
\end{bmatrix}$ as $\text{MSE}_k$, which can be calculated as
\begin{align}
    \text{MSE}_k&=\mathbb{E}_{Q_k}\left[ \frac{1}{T}\int_{(k-1)T+d}^{kT+d} \mbox{MSE}(t)\mathrm{d}t\right] \\
    &=\frac{1}{T}\int_{d}^{T+d} \left( \sigma^2t+\mathbb{E}\left[Q_k^2\right] \right)\mathrm{d}t \\
    &=\frac{\sigma^2T}{2}+\sigma^2d+\mathbb{E}[Q_k^2].
\end{align}

Finally, in an infinite horizon, the average MSE can be calculated over all periods as
\begin{align}
    \text{MSE}&= \limsup_{K\to \infty}\dfrac{T\sum_{k=1}^K \text{MSE}_k+\int_0^{T+d} \text{MSE}(t)\mathrm{d}t }{KT}, \\
    &=\limsup_{K\to \infty} \dfrac{K\frac{T^2}{2}+T\sum_{k=1}^K\mathbb{E}[Q_k^2]+KT\sigma^2d+\frac{(T+d)^2}{2} }{KT}\\
    &=\sigma^2\frac{T}{2}+\sigma^2d+\mathbb{E}[Q^2], \label{eq:mse_mean}
\end{align}
where $\mathbb{E}[Q^2]$ stands for $\limsup_{K\to\infty}\frac{1}{K}\sum_{k=1}^K\mathbb{E}[Q_k^2]$ which converges to a finite limit under the assumption that the second moment of the quantization error is uniformly bounded. This assumption is required to obtain a finite MSE, and is verified for our proposed quantization methods in Section~\ref{sec:results}.

\section{Optimum Quantization}
The necessary conditions for an MSE-optimum quantizer for the input variable $Y_k$ listed in \cite{gersho2012vector} are: i) The conditional expectation of each region should be the center point, i.e., $\mathbb{E}[Y_k|Y_k\in\mathcal{R}_{kn}]=c_{kn}$. ii) The region boundary should satisfy the nearest neighborhood condition, i.e., $|Y_k-c_{km}|\geq |Y_k-c_{kn}|$ for $Y_k \in \mathcal{R}_{kn}$, $n,m\in\mathcal{C}$, $n\neq m$. From the second condition, the boundaries of the quantization regions  for our case can be expressed as
\begin{align}
    \tau_{k-}=\frac{c_{k-}+c_{k\emptyset}}{2}, \ \tau_{k+}=\frac{c_{k+}+c_{k\emptyset}}{2}, \ \forall k. 
\end{align}

The well-known Lloyd-Max algorithm is an iterative algorithm that starts from an arbitrary quantization function and finds a better quantization function by satisfying the aforementioned conditions until it converges \cite{gersho2012vector}, and it is described for our problem in Algorithm \ref{alg:lloyd1}. The algorithm converges to an optimum quantization function if the input distribution is a log-concave function \cite{gersho2012vector}.

\begin{algorithm}[h]
    \caption{Lloyd Quantization}\label{alg:lloyd1}
    \begin{algorithmic}
        \State \textbf{Initialize:}  $\tau_{-}^*=F_Y^{-1}(1/3)$, $\tau_{+}^*=F_Y^{-1}(2/3)$.
                 \While{$|\tau_--\tau_-^*| > \epsilon$ OR $|\tau_+-\tau_+^*|>\epsilon$  }
         \State{Update thresholds:} $\tau_{-} \gets \tau_{-}^*$, $\tau_{+} \gets \tau_{+}^*$ 
         \State{Obtain new center points:} $c_-=\dfrac{\int_{-\infty}^{\tau_-}yf_Y(y)\mathrm{d}y}{\int_{-\infty}^{\tau_-}f_Y(y)\mathrm{d}y}$, $c_\emptyset=\dfrac{\int_{\tau_-}^{\tau_+}yf_Y(y)\mathrm{d}y}{\int_{\tau_-}^{\tau_+}f_Y(y)\mathrm{d}y}$, $c_+=\dfrac{\int_{\tau_+}^{\infty}yf_Y(y)\mathrm{d}y}{\int_{\tau_+}^{\infty}f_Y(y)\mathrm{d}y}$. 
         \State{Obtain new thresholds:}
         $\tau_-^*\gets \dfrac{c_-+c_\emptyset}{2}$, $\tau_+^*\gets \dfrac{c_\emptyset+c_+}{2}$.
        \EndWhile
    \end{algorithmic}
\end{algorithm}

\begin{figure}[t]
    \centering
    \includegraphics[width=0.95\linewidth]{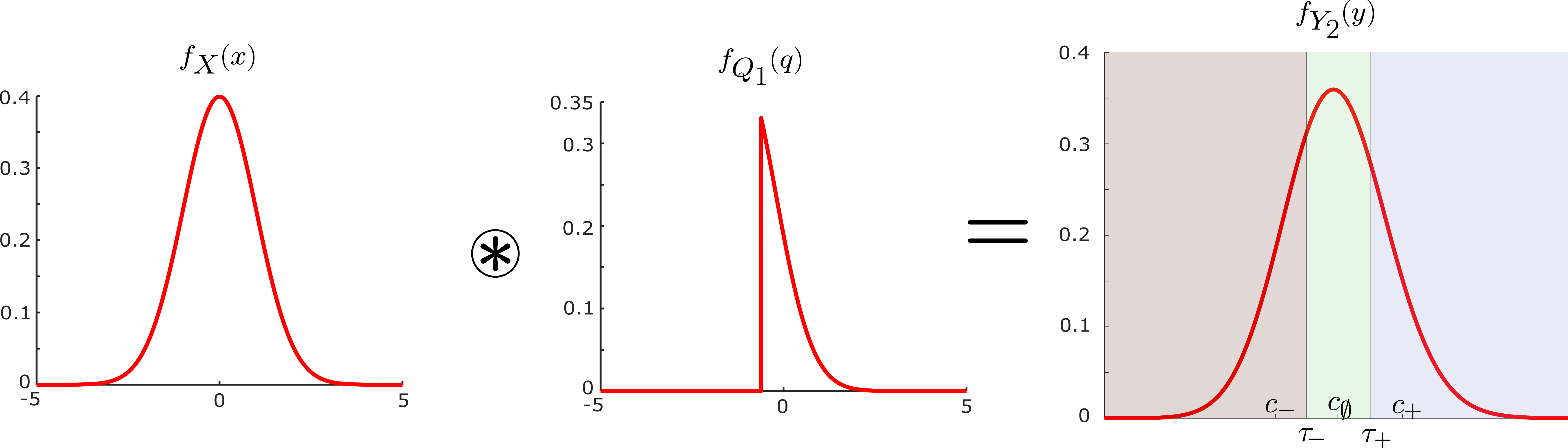}
    \caption{Obtaining $f_{Y_2}(y)$ from $f_{Q_1}(q)$ when the corresponding bit for the symbol `$+$' is transmitted in the first period.}
    \label{fig:conv_p}
\end{figure}

As described with the system model, the exact value of the quantization error $Q_k$ is only available to the source. On the other hand, the monitor can calculate the distribution of $Q_k$ from the distribution of $Y_k$. Let us consider a quantization function $\Phi_k$ and the quantized variable $Y_k \in \mathcal{R}_{kn}$ with a known distribution $f_{Y_k}(y)$. The distribution of the quantization error $f_{Q_k}(q)$ is the shifted, truncated, and normalized version of the $f_{Y_k}(y)$, and can be calculated as
\begin{align}
    f_{Q_k}(q)=\begin{cases}
        \dfrac{f_{Y_k}(q+c_n)}{\int_{R_{kn}}f_{Y_k}(y)\mathrm{d}y}, & q+c_n \in \mathcal{R}_{kn}, \\ 
        0, & \text{otherwise}.
    \end{cases}
\end{align}
Recursively, the distribution of the next input $Y_{k+1}=X_{k+1}+Q_{k}$ can be obtained by the convolution as $f_{Y_{k+1}}(y)=f_X(y) * f_{Q_{k}}(y)$. An application of this procedure is illustrated in Fig.~\ref{fig:conv_p}.

Note that in the first iteration, the variable quantized $Y_1=X_1$ has a normal distribution, which is log-concave. In the second iteration, the variable quantized is $Y_2=X_2+Q_2$, where the distribution of $Q_2$ is a shifted and truncated version of the distribution of $Y_1$, thus, it is also log-concave. Additionally, the distribution of $Y_2$ is obtained by the convolution of two log-concave functions, which preserves the log-concavity \cite{dharmadhikari1988unimodality}. Thus, by induction, the Lloyd-Max algorithm converges to the optimum quantization functions for our setting.

\begin{algorithm}[t]
    \caption{Optimum Quantization}\label{alg:quant}
    \begin{algorithmic}
        \State \textbf{Initialize:}  $f_{Y_1}(y)=f_X(y)$, $Q_0=0$, and distribution of quantization error is
        \begin{align}
            f_{Q_0}(q)=\begin{cases}
                1, & q=0, \\ 0, & o.w.
            \end{cases}
        \end{align}
                 \For{$K=1\dots $ }
         \State\textbf{Step 1 (New Sample)}: Generate a new sample $X_k\sim\mathcal{N}(0,\sigma^2T)$, and $Y_k=X_k+Q_{k-1}$.
        \State \textbf{Step 2 (Lloyd Quantization)}: Obtain new quantizater $\Phi_k$ by Algorithm~\ref{alg:lloyd1} for $f_{Y_k}(y)$.
        \State\textbf{Step 3 (Apply Quantization)}: Apply quantization to the input $Y_k=X_k+Q_{k-1}$, and obtain quantization error as
        \begin{align}
            Q_k=y_k-c_n, \ Y_k\in \mathcal{R}_n. 
        \end{align}
        \State\textbf{Step 4 (Obtain distributions)}: Obtain distribution of quantization error and new input
        \begin{align}
            f_{Q_k}(q)&=\begin{cases}
                \dfrac{f_{Y_{k}}(q+c_n) }{\int_{R_n} f_{Y_k}(y)\mathrm{d}y}, & q+c_n\in R_n, \\ 0, & o.w.,
            \end{cases}\\
            f_{Y_{k+1}}(y)&= f_{X}(y) * f_{Q_{k}}(y).
        \end{align}
        \EndFor
    \end{algorithmic}
\end{algorithm}

We propose a quantization method in which the distributions of variables $Y_k$ and $Q_k$ are tracked by the source and the monitor, and an optimum quantization function is obtained for each period $k$. The whole procedure of this quantization method is summarized in Algorithm~\ref{alg:quant}.

\section{Numerical Results}\label{sec:results}

\begin{table*}[t]
    \centering
    \caption{Parameters for different quantization methods}
    \label{tab:quant}
    \begin{tabular}{ccccc}
    & \multicolumn{3}{c}{last bit aware}  &   Gaussian approximation   \\ \cline{2-5} 
    \multicolumn{1}{c|}{}                & \multicolumn{1}{c|}{$\Phi_-$}      & \multicolumn{1}{c|}{$\Phi_\emptyset$} & \multicolumn{1}{c|}{$\Phi_+$}     & \multicolumn{1}{c|}{$\Phi_G$} \\ \cline{2-5} 
    \multicolumn{1}{c|}{$c_-$}           & \multicolumn{1}{c|}{$-1.489\sigma\sqrt{T}$} & \multicolumn{1}{c|}{$-0.048\sigma\sqrt{T}$}     & \multicolumn{1}{c|}{$1.326\sigma\sqrt{T}$} & \multicolumn{1}{c|}{$-1.360\sigma\sqrt{T}$} \\ \cline{2-5} 
    \multicolumn{1}{c|}{$c_{\emptyset}$} & \multicolumn{1}{c|}{$-1.310\sigma\sqrt{T}$} & \multicolumn{1}{c|}{$0$}         & \multicolumn{1}{c|}{$1.310\sigma\sqrt{T}$} & \multicolumn{1}{c|}{$0$}      \\ \cline{2-5} 
    \multicolumn{1}{c|}{$c_{+}$}         & \multicolumn{1}{c|}{$-1.326\sigma\sqrt{T}$} & \multicolumn{1}{c|}{$0.048\sigma\sqrt{T}$}      & \multicolumn{1}{c|}{$1.489\sigma\sqrt{T}$} & \multicolumn{1}{c|}{$1.360\sigma\sqrt{T}$}  \\ \cline{2-5} 
    \multicolumn{1}{l|}{$\tau_{-}$}      & \multicolumn{1}{c|}{$-0.768\sigma\sqrt{T}$}  & \multicolumn{1}{c|}{$-0.655\sigma\sqrt{T}$}     & \multicolumn{1}{c|}{$-0.639\sigma\sqrt{T}$} & \multicolumn{1}{c|}{$-0.680\sigma\sqrt{T}$}  \\ \cline{2-5} 
    \multicolumn{1}{l|}{$\tau_{+}$}      & \multicolumn{1}{c|}{$0.639\sigma\sqrt{T}$}   & \multicolumn{1}{c|}{$0.655\sigma\sqrt{T}$}      & \multicolumn{1}{c|}{$0.768\sigma\sqrt{T}$}  & \multicolumn{1}{c|}{$0.680\sigma\sqrt{T}$}   \\ \cline{2-5} 
    \end{tabular} 
\end{table*}

For optimum quantization functions, we set the tolerance value to $\epsilon=1\cdot10^{-4}$, and discretize the distributions over $\begin{bmatrix}
    -7 \sigma\sqrt{T}, & 7\sigma\sqrt{T}
\end{bmatrix}$ with the step size $\mathrm{d}t=1\cdot10^{-3}$. With this method, we observe that the average distortion converges to $\lim_{K\to\infty}\mathbb{E}[Q_K^2]=\mathbb{E}[Q^2]=0.2367\sigma^2T$. However, obtaining a new quantization function in each step from Algorithm~\ref{alg:lloyd1} may be time-consuming, and may not be feasible for sensors with limited computational resources. Therefore, we also propose two alternative quantization methods: \emph{last-bit aware} quantization and \emph{Gauassian aproximation} based quantization.

\paragraph{Last-Bit Aware Quantization}
In this quantization method, we consider three predefined quantization functions $\Phi_-$, $\Phi_\emptyset$, and $\Phi_+$ that are available to both the encoder and the decoder. Following the sampling/receiving a symbol $n\in\mathcal{C}$, the encoder/decoder alters its quantization functions to $\Phi_n$. For this scheme, we first run the optimum quantization functions over $10^6$ iterations, then average the parameters based on the last-transmitted bit. The result obtained from this method are shown in Table~\ref{tab:quant}. For instance, the column $\Phi_+$ refers to the average of the optimum quantization parameters following the bit that corresponds to the symbol `$+$'  being transmitted.

\paragraph{Gaussian Approximation Based Quantization}
In this quantization, we assume that the quantization error $Q_k$ and the quantized variable $Y_k$ are Gaussian distributed at each period. This approximation is valid for the first input $Y_1$, where the optimum quantization function gives $\mathbb{E}[Q_1^2]=0.1902\sigma^2T$. If the quantization error is assumed to be Gaussian, in the next step, a Gaussian input, $Y_2$, with $\mathbb{E}[Y_2^2]=1.1902\sigma^2T$ will be quantized, which will again give us $\mathbb{E}[Q_2^2]=0.1902\cdot1.1902\sigma^2T$. By geometric series summation, we have $\lim_{K\to\infty}\mathbb{E}[Q_K^2]=\mathbb{E}[Q^2]=0.2349\sigma^2T$. It is not surprising that this distortion value is close to that obtained from the optimum quantization method. Indeed, when we empirically compare the distribution of $Y_k$ with a Gaussian distribution $f_{\hat{Y}_k}(y)$ which has the same variance as $Y_k$ by using relative entropy (KL distance) \cite{cover1999elements} over their discretized distributions, we observe that the average of the relative entropy is lower than $10^{-3}$, and it becomes almost zero if the empty bit is transmitted. From this approximation, we propose the following quantization method. We consider a static quantization function that is used in each step, which is obtained for the Gaussian distributions with $\sigma_G^2=1.2349\sigma^2T$. The parameters of the obtained quantization function are given in Table~\ref{tab:quant} under the column of $\Phi_G$.

Up to this point, we have considered a deterministic channel delay $d$ that allows the monitor to know when to expect to receive a bit. Therefore, when the input falls into the region $\mathcal{R}_{\emptyset}$, the monitor updates its estimation with $c_{\emptyset}$. However, when there is a random delay, this method will not be applicable. On the other hand, because the center point for $\mathcal{R}_\emptyset$ is $c_\emptyset=0$ for $\Phi_G$ for every period, the receiver does not need to exactly know when it is supposed to receive an update. Furthermore, MSE expression in \eqref{eq:mse_mean} can be adapted as
\begin{align}
    \text{MSE}=\sigma^2\frac{T}{2}+\mathbb{E}[Q^2]+\sigma^2\mathbb{E}[D], \label{eq:mse_del}
\end{align}
for a random delay $D$ that is guaranteed to be always smaller than the period $T$, i.e., $\mathbb{P}[D>T]=0$. For a random delay with $\mathbb{P}[D>T]>0$, queueing and preemption policies should be considered, which are not considered in this paper.

Another observation is that, for all of the quantization methods we have considered, the quantization functions' output is not equally probable. In addition, the source does not transmit any bits when the input is in the region $\mathcal{R}_{\emptyset}$, with no cost. Therefore, we define the transmission rate (TR) which corresponds to how many times an actual bit is transmitted per unit time. TR can be calculated as 
\begin{align}
    \text{TR}=\dfrac{1-p_{\emptyset}}{T},
\end{align} where $p_\emptyset$ is the probability of transmitting the empty bit.

Table~\ref{tab:sampling} shows the expected distortion and transmission rates for different methods. Notice that the transmission rate only depends on the structure of the quantization function, and the expected distortion does not depend on the delay. We observe that all methods result in similar performance, since quantized variables $Y_k$ are almost Gaussian as discussed.

\begin{table}[]
    \caption{Comparison of the expected distortion and sampling rates of different quantization methods}
    \begin{tabular}{lll}
    & \multicolumn{1}{c}{$\mathbb{E}[Q^2]$}  & \multicolumn{1}{c}{TR}               \\ \cline{2-3}
    \multicolumn{1}{l|}{Optimum Quantization}      & \multicolumn{1}{l|}{$0.2367 \sigma^2T$}  & \multicolumn{1}{l|}{$0.54/T$} \\ \cline{2-3} 
    \multicolumn{1}{l|}{Last-bit Aware}         & \multicolumn{1}{l|}{$0.2369 \sigma^2T$}  & \multicolumn{1}{l|}{$0.54/T$} \\ \cline{2-3} 
    \multicolumn{1}{l|}{Gaussian Approximation} & \multicolumn{1}{l|}{$0.2390  \sigma^2T$} & \multicolumn{1}{l|}{$0.54/T$} \\ \cline{2-3} 
    \end{tabular} 
    \label{tab:sampling}
\end{table}

\begin{figure}[t]
    \centering
    \includegraphics[width=0.99\linewidth]{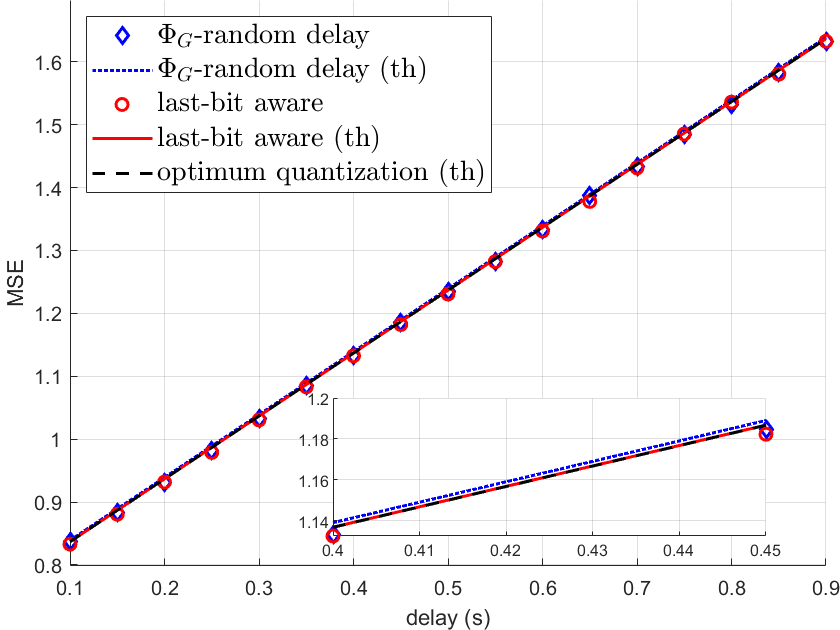}
    \caption{MSE comparison of different quantization functions. Lines correspond to the analytical results obtained from \eqref{eq:mse_gen} using average distortions in Table~\ref{tab:quant}.}
    \label{fig:sim_res}
\end{figure}

To compare MSE performances of different quantization methods, we fix the parameters $\sigma^2=1$ and $T=1$, and consider a varying delay. For random delay, we consider a uniform random delay in the interval $\begin{bmatrix} d-0.05 & d+0.05    \end{bmatrix}$. The results are shown in Fig.~\ref{fig:sim_res}, where lines correspond to the theoretical results obtained from \eqref{eq:mse_mean} and \eqref{eq:mse_del} for the corresponding $\mathbb{E}[Q^2]$ from Table~\ref{tab:sampling}. These simulation results verify our analytical approach for both deterministic and random delay cases.

\bibliographystyle{unsrt}
\bibliography{bibl}
\end{document}